\documentclass[twocolumn,aps,showpacs,preprintnumbers,amsmath,amssymb]{revtex4}
\usepackage[dvips]{hyperref}
\usepackage{graphicx}
\usepackage{dcolumn}
\usepackage{bm}

\newcommand{\ccc}{Cs$_2$CuCl$_4$ }
\newcommand{\tcc}{TlCuCl$_3$ }

\begin{document}

\title{Bose-Einstein Condensation of Magnons in \ccc}

\author{T. Radu$^1$, H. Wilhelm$^1$, V. Yushankhai$^{1,2}$,
  D. Kovrizhin$^3$, R. Coldea$^4$, Z. Tylczynski$^{5}$, T. L\"uhmann$^1$, and F. Steglich$^1$}
\affiliation{$^1$Max--Planck--Institut f\"ur Chemische Physik
fester Stoffe, N\"othnitzer Str. 40, 01187 Dresden, Germany\\
\noindent $^2$Laboratory of Theoretical Physics,
JINR, 141980 Dubna, Russia\\
\noindent $^3$Max--Planck--Institut f\"ur Physik komplexer
Systeme, N\"othnitzer Str. 38, 01187 Dresden, Germany\\ 
\noindent$^4$Oxford Physics, Clarendon Laboratory, Parks Road, Oxford OX1
3PU, Great Britain\\ 
\noindent$^5$Institute of Physics, Adam Mickiewicz University, Umultowska
85, 61-614 Poznan, Poland}

%
%
\begin{abstract}
  We report on results of specific heat measurements on single
  crystals of the frustrated quasi--2D spin--$1/2$ antiferromagnet
  Cs$_2$CuCl$_4$ ($T_N=0.595$~K) in external magnetic fields $B<12$~T
  and for temperatures $T>30$~mK. Decreasing $B$ from high fields
  leads to the closure of the field-induced gap in the magnon spectrum
  at a critical field $B_c\simeq 8.51$~T and a magnetic phase
  transition is clearly seen below $B_c$. In the vicinity to $B_c$,
  the phase transition boundary is well described by the power-law
  $T_c(B)\propto (B_c-B)^{1/\phi}$ with the measured critical exponent
  $\phi\simeq 1.5$. These findings are interpreted as a Bose-Einstein
  condensation of magnons.
\end{abstract}


\maketitle

%
%
In a quantum antiferromagnet (AFM) a fully spin-polarized state can be
reached at high magnetic field $B$ exceeding a saturation field $B_c$.
In this state, spin excitations are gapped {\it ferromagnetic}
magnons. With decreasing $B$ and passing through $B_c$, an
antiferromagnetic long-range order of the transverse spin component
develops. Provided the symmetry of the spin Hamiltonian is such that
the rotational invariance around the applied field is preserved, the
transverse spin component ordering can be regarded as a Bose-Einstein
condensation (BEC) in a dilute gas of magnons. This concept was
formulated theoretically many years ago \cite{Matsubara56,Batyev84}.
For most of the known AFMs, $B_c$ can be well above 100~T. An
exceptionally low and easily accessible saturation field of
$B_c\approx 8.5$~T, however, is needed in the quantum spin-1/2 AFM
Cs$_2$CuCl$_4$. In this system the dominant exchange spin coupling $J$
is rather weak, $J=4.34(6)$~K \cite{Coldea96}. The other isotropic
spin coupling constants and the anisotropic Dzyaloshinsky-Moriya (DM)
interaction are smaller and were determined with high accuracy by
neutron experiments \cite{Coldea02}. Thus, the spin Hamiltonian
involves the isotropic exchange $H_{0}$, the DM anisotropic term
$H_{DM}$ and the Zeeman energy $H_{B}$ and is given by $H = H_0 +
H_{DM} + H_{B}$.

\ccc falls into the class of easy-plane AFMs with
$U(1)$-rotational invariance around the crystallographic $a$-axis.
Thus, for $B$ applied along the $a$-axis, the $U(1)$ symmetry can be
broken spontaneously due to the transverse spin component ordering at
$T_c$. This is accompanied by the appearance of a Goldstone mode with
linear dispersion, which is interpreted as signature of a magnon BEC
\cite{Coldea02}. However, an unambiguous evidence for a BEC
description of the field-induced phase transition would be the
determination of the critical exponent $\phi$ in the field dependence
of the critical temperature
\begin{equation}\label{eq:tcvsb}
  T_c(B)\propto (B_c-B)^{1/\phi}~.
\end{equation}
Theory for a 3D Bose gas predicts a universal value $\phi_{\rm
  BEC}=3/2$ \cite{Nohadani04}, which coincides with the result of a
mean-field treatment \cite{Nikuni00}.

A magnon BEC in \tcc was recently reported \cite{Nikuni00,Rueegg03}.
In this quantum AFM with a dimerized spin-liquid ground state, the
saturation field is rather high, $B_{c,2}\approx 60$~T, and the BEC
transition was studied near the first critical field, $B_{c,1}\simeq
5.6$~T. At $B=B_{c,1}$, the singlet-triplet excitation gap is expected
to close and a BEC occurs for $B>B_{c,1}$ \cite{Nikuni00}. However, a
few experimental findings show deviations from a pure magnon BEC: An
anisotropic spin coupling (of unknown nature) might produce a small
but finite spin gap in the ordered state for $B>B_{c,1}$
\cite{Kolezhuk04} and the reported critical exponent $\phi$ is
somewhat larger than predicted by theory \cite{Nikuni00}.

In this Letter we report on specific heat measurements
\cite{Wilhelm04} on single crystals of \ccc at low temperatures
(30~mK~$<T<6$~K) and high magnetic fields ($B<12$~T) applied along
the crystallographic $a$-axis. The aim of this thermodynamic study
was (i) to trace the field dependence of $T_c(B)$ near $B_c$,
i.e., to extract the power law according to eq.~\ref{eq:tcvsb},
and (ii) to determine the closure of the spin gap. The access to
very low temperatures ($T/J\simeq 10^{-2}$) enabled us to be as
close as possible to the asymptotic regime where universal scaling
laws are expected to hold.

%
%

\begin{figure}
\center{\includegraphics[width=0.90\columnwidth,clip]{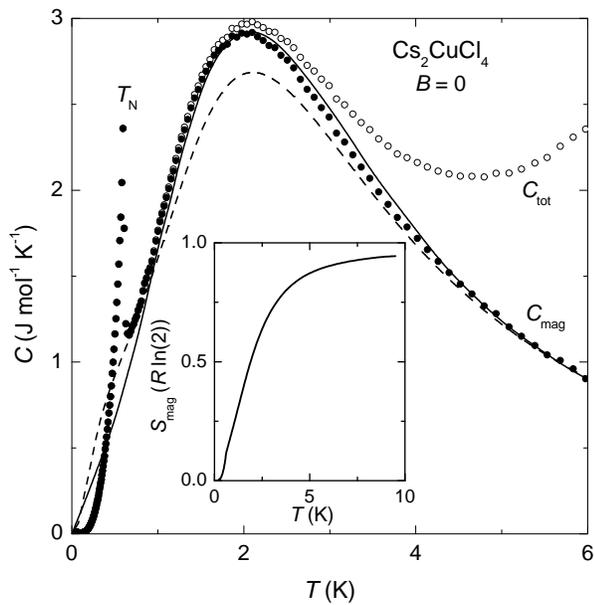}}
 \caption{Specific heat of \ccc in zero magnetic field. From the total
   specific heat, $C_{tot}$ (open symbols) the phonon contribution
   $C_{ph}$ has been subtracted to obtain the magnetic part $C_{mag}$
   (filled symbols). The solid and dashed lines represent the calculated $C(T)$
   (see text). Inset: Temperature dependence of the magnetic entropy, $S_{mag}(T)$.}
 \label{fig:zerofield}
\end{figure}

Figure~\ref{fig:zerofield} shows the specific heat of \ccc in zero
magnetic field. The magnetic contribution, $C_{mag}$ to the total
specific heat, $C_{tot}$, was obtained by subtracting the phonon
contribution
$C_{ph}=13599\left(T/\Theta_D\right)^3$~Jmol$^{-1}$K$^{-1}$ (using a
Debye temperature $\Theta_D=126$~K). Furthermore, the contribution of
the nuclear specific heat to the total specific heat at very low
temperatures has been accounted for in the analysis of all the raw
data shown in the following \cite{alphab2,NMR}. The two prominent
features present in $C(T)$ are the broad maximum related to the
cross-over from the paramagnetic to a short-range spin correlated
state and the $\lambda$-like peak. The latter is the signature of the
entrance into the 3D magnetically ordered state ($T_N=0.595$~K), where
the magnetic structure is a spiral in the ($b,c$) plane
\cite{Coldea96}. The overall shape of $C(T)$ above $T_N$ is already
captured quantitatively by including only the strongest term in the
Hamiltonian, the dominant coupling $J$ (solid line) \cite{Kluemper98}.
Exact theoretical predictions are not available for the full
Hamiltonian in Cs$_2$CuCl$_4$, but including the next order term, the
frustrated 2D zig-zag coupling $J^{\prime}=J/3$, and using a
high-temperature series expansion technique \cite{Weihong}, give
similar behavior (dashed line) although with a slightly lower specific
heat at the position of the broad maximum. It may be possible that
including the other small terms like the DM interaction ($D_a/J
\approx 5\%$) could improve the agreement. The magnetic entropy
$S_{mag}(T)$ shows that $S= R\ln 2$ for spin-1/2 is almost recovered
at 10~K (inset to Fig.~\ref{fig:zerofield}) which corresponds roughly
to the observed band width of the spin excitations at base
temperature \cite{Coldea01}.

%
%

\begin{figure}
\center{\includegraphics[width=1\columnwidth,clip]{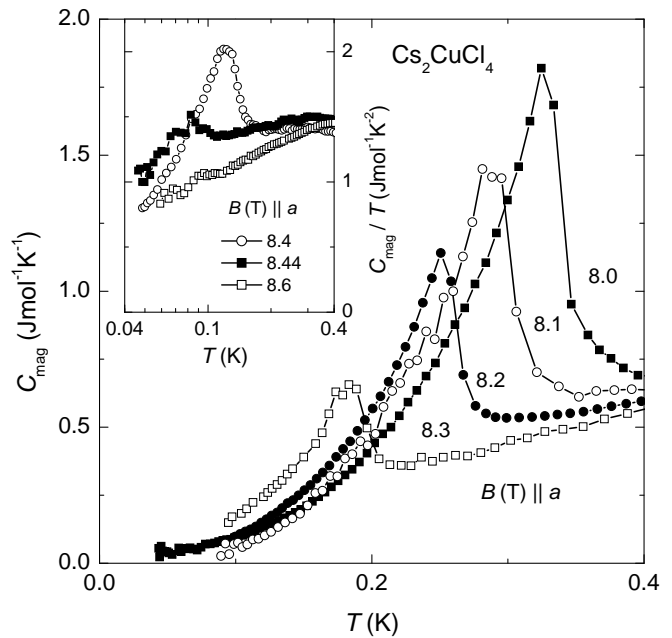}}
\caption{Magnetic specific heat, $C_{mag}$ vs $T$ of \ccc close to the critical field.
  Inset: $C_{mag}/T$ vs $T$ in a semi-logarithmic plot. The data
  recorded at $B=8.44$~T reveal a small jump at $T_c=76$~mK,
  indicating the vicinity of the critical field.}
\label{fig:closebc}
\end{figure}

The ordering temperature and the position of the broad maximum hardly
change for small fields. However, the transition temperature to the
spiral ordered state, which can be regarded as a cone-like structure
\cite{Coldea02}, varies very strongly above 8 T
(Fig.~\ref{fig:closebc}). Instead of $T_N$, we label the transition
temperature in this field range $T_c$, as in eq.~\ref{eq:tcvsb}, in
order to follow the nomenclature used in the theoretical description.
The $\lambda$-like anomaly in $C_{mag}(T)$ is gradually suppressed in
its height and its position is pushed to lower temperatures with
increasing field. An extraordinary change occurs as the field is
increased from 8.4~T to 8.44~T (inset to Fig.~\ref{fig:closebc}).
Upon this tiny field change ($\Delta B/B<0.5\%$), $T_c$ is reduced by
almost a factor of two ($T_c=76$~mK at $B=8.44$~T), and $T_c$ has
shifted downwards by almost one order of magnitude compared to the
zero-field value. No further signatures of the transition can be
resolved in our data upon approaching the critical field $B_c\simeq
8.5$~T. For $B>B_c$ the ordering of the transverse component of the
magnetic moment completely disappears since the spin system enters a
field-induced ferromagnetic (FM) state \cite{Coldea02}.

%
%

\begin{figure}
\center{\includegraphics[width=1\columnwidth,clip]{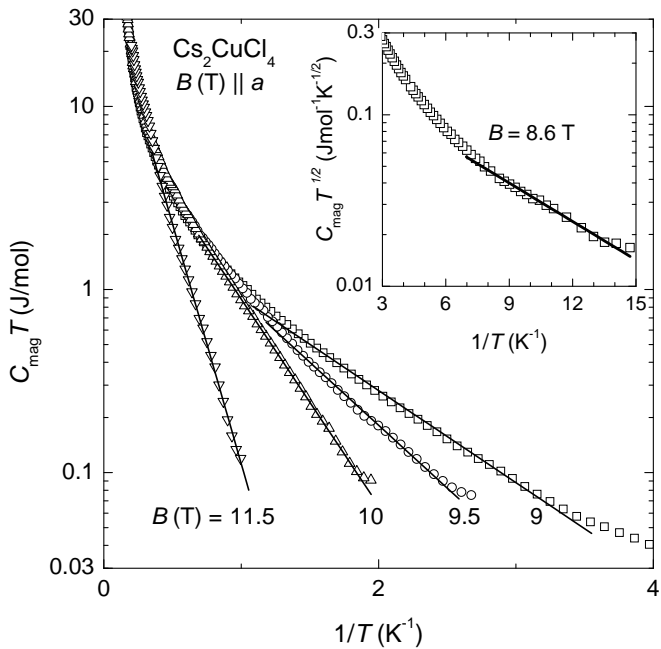}}
 \caption{Semi-logarithmic plot of $C_{mag} T$ vs $1/T$ of \ccc for
   fields above $B_c$. In this representation the slope of the data
   (solid lines) yields the value of the gap $\Delta$ present in the magnon
   excitation spectrum. The data shown in the inset were obtained at
   $B=8.6$~T and are plotted as $C_{mag}\sqrt T$ vs $1/T$.}
\label{fig:abovebc}
\end{figure}

A field-induced gap in the magnon excitation spectrum was observed in
the FM state by neutron scattering measurements \cite{Coldea02}. Its
field dependence was given to be $\Delta=g\mu_B(B-B_c)$, with $B_c
\simeq 8.44$~T, $g=2.18$, and $\mu_B$ the Bohr magneton. For the
interpretation of the phase transition below $B_c$ as a BEC of magnons
it is crucial that the gap closes at $B_c$. To provide a compelling
evidence for this fact we re-examined the phase diagram
\cite{Coldea02} above $B_c$ with our thermodynamic measurements.
The results are presented in Fig.~\ref{fig:abovebc}.

The magnon dispersion along the $a$-direction is small due to a
weak interlayer spin coupling. Thus, for temperatures well above a
characteristic energy scale $E^{\ast}\approx 50$~mK, the actual
magnon dispersion is of 2D character. However, for $T\lesssim
E^{\ast}$ a smooth cross-over to a 3D character is expected.
Assuming first a 2D quadratic magnon dispersion, the leading
contribution to the temperature dependence of the specific heat is
given by $C_{mag}\simeq \exp(-\Delta/T)/T$, provided that
$T<\Delta$. As shown in Fig.~\ref{fig:abovebc}, this behavior fits
well the experimental data above 0.3~K. The obtained field dependence
of $\Delta(B)$ is discussed below. The deviation from a straight line of the
9~T data below 0.3~K in Fig.~\ref{fig:abovebc} might indicate the
cross-over from 2D to 3D magnons. This notion is supported by the
low-temperature data plotted as $C_{mag}\sqrt T$ vs $1/T$ in the
inset to Fig.~\ref{fig:abovebc}. This presentation is used since a
3D dispersion relation yields as leading term in the specific heat
$C_{mag}\simeq \exp(-\Delta/T)/\sqrt T$. The straight line below
$\approx 0.15$~K indicates that this model describes the data
appropriately. However, in this cross-over region, the 2D model
describes the data equally well and the determined value of the
gap is (within the error bar) the same as the one deduced from the
$C_{mag}T$ vs $1/T$ plot.

%
%

\begin{figure}
\center{\includegraphics[width=1\columnwidth,clip]{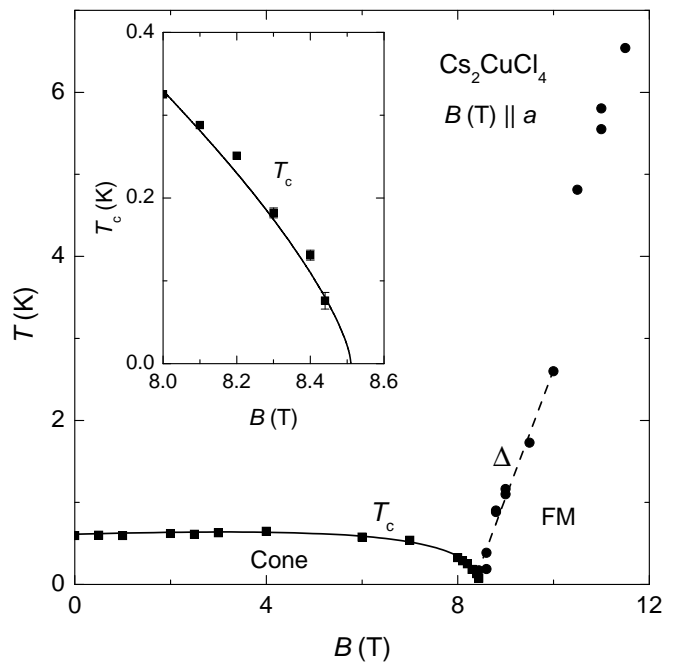}}
\caption{($T,B$) phase diagram of \ccc for $B\parallel a$. The
  ordering temperature $T_c$ decreases as the magnetic field
  approaches the critical field $B_c = 8.51$~T. Above $B_c$ a field-polarized
  ferromagnetic (FM) state is entered and the gap $\Delta$ in the spin
  excitation spectrum opens. Inset: The experimental $T_c(B)$ data
  points are well described by the calculated phase boundary of the
  BEC of magnons (solid line).}
\label{fig:phasediagram}
\end{figure}

The ($T,B$) phase diagram of \ccc obtained from our specific heat
experiments is presented in Fig.~\ref{fig:phasediagram}. The field
dependence of the magnetic transition temperature is in very good
agreement with $T_c(B)$ obtained from neutron data up to 8~T
\cite{Coldea01}. The specific heat data, however, revealed that $T_c$
starts to decrease strongly above 8~T and $T_c\rightarrow 0$ for
$B\rightarrow B_c$. Fitting the power law dependence $T_c(B)\propto
(B_c-B)^{1/\phi}$ to the data for $B\geq 8$~T with the assumption of
$B_c=8.50$~T yields an exponent $\phi=1.52(10)$. We want to stress
that the value of $\phi$ is very sensitive to the chosen value of
$B_c$. An exponent $\phi=1.44(10)$ is obtained if $B_c=8.51$ is used.
The solid line plotted in the inset to Fig.~\ref{fig:phasediagram}
represents the result of the theoretical analysis described below.
Above $B_c$ the fully spin-polarized FM state is created and the gap $\Delta$
opens in the spin excitation spectrum. The dashed line represents a
linear fit to the data for $B\leq 10$~T (main part of
Fig.~\ref{fig:phasediagram}). This yields $B_c=8.3(10)$~T and
$g=2.31(15)$. The relatively large errors are due to the uncertainties
in the fit.

%
%

%
%

To treat the observed phase transition slightly below the saturation
field $B_{c}$ as a BEC of magnons \cite{Matsubara56,Batyev84,Nikuni00}
we used the hard-core boson representation for spin-$1/2$ operators
$S_{i}^{\pm},S_{i}^{z}$ in the original Hamiltonian $H$. Because the
DM interaction ($D=0.053(5)J$ \cite{Coldea02}) changes sign between
even and odd magnetic layers, which are stacked along the
$a$-direction, two types of bosons, $a_{i}$ and $b_{j}$ are introduced
for the two types of layers \cite{mapping,Veillette05}. The hard-core boson
constraint was satisfied by adding to $H$ an infinite on-site
repulsion, $U\to\infty$, between bosons given by

\begin{equation}
H_{U}^{\left(  a\right)  }+H_{U}^{\left(  b\right)
}=U\sum_{i}a_{i}^{+}
a_{i}^{+}a_{i}a_{i}+U\sum_{j}b_{j}^{+}b_{j}^{+}b_{j}b_{j}.\label{theory1}
\end{equation}

The interlayer coupling $J''=0.045(5)J$ \cite{Coldea02} mixes $a$ and
$b$ boson modes and results in two bare magnon excitation branches $A$
and $B$. Their dispersion relations are given by \cite{Coldea02}

\begin{equation}
E_{q}^{A,B}=J_{q}\mp\operatorname{sign}D_{q}\sqrt{D_{q}^{2}+\left( J_{q}^{^{\prime\prime}}\right)^{2}}-E_{0},
\label{theory2}
\end{equation}

\noindent
with
\begin{eqnarray}
  J_{q} &=& J\cos q_{x}+2J^{\prime}\cos\left( q_{x}/2\right) \cos\left(q_{y}/2\right)~, \\
  D_{q} &=& 2D\sin\left( q_{x}/2\right) \cos\left(q_{y}/2\right)~, \\
  J_{q}^{^{\prime\prime}} &=&
  J^{\prime\prime}\cos\left(q_{z}/2\right)~.
\end{eqnarray}

\noindent Here $J'=0.34(3)J$ \cite{Coldea02} and the $q$-values
are restricted to $0\leqslant q_{x}<2\pi$, $0\leqslant
q_{y}<4\pi$, and $0\leqslant q_{z}<2\pi$.

The degenerate minima
$E_{\vec{Q_1}}^{A}=E_{\vec{Q_2}}^{B}=0$
are at $\vec{Q}_{1}=\left(\pi+\delta_{1},0,0\right)$ for branch
$A$ and at $\vec{Q}_{2}=\left(\pi-\delta_{2},2\pi,0\right)$ for
branch $B$. Without loosing precision we can use
$\delta_{1}\simeq\delta_{2}\simeq\delta=2\arcsin\left(J^{\prime}/2J\right)$.
The bilinear part of $H$ now reads

\begin{equation}
H_{bil}=\sum_{q}\left[  \left(  E_{q}^{A}-\mu_{0}\right)  A_{q}^{+} A_{q}+\left(  E_{q}^{B}-\mu_{0}\right)  B_{q}^{+}B_{q}\right],
\label{theory3}
\end{equation}

\noindent
with $A_{q}=\alpha_{q}a_{q}+\beta_{q}b_{q}$,
$B_{q}=\alpha_{q}b_{q}-\beta_{q}a_{q}$,
$\alpha_{q}^{2}+\beta_{q}^{2}=1$, and the bare chemical potential
$\mu_{0}=g\mu_{B}\left(B_{c}-B\right)$. The saturation field
$B_{c}=W/(g\mu_{B})$, with $W$ being the magnon bandwidth, was
calculated to be $B_{c}=8.51$~T assuming $g=2.20$ \cite{Veillette05,Bailleul94}.

The interaction given by eq.~\ref{theory1} describes the
scattering of $A$ and $B$ magnons. Near the quantum critical
point, $\left(B_{c}-B\right)\ll B_{c}$ and at low temperature, the
average density of magnons $n^{A}=n^{B}=n$ is low, $n\sim\left(
1-B/B_{c}\right)$. The magnon scattering can be treated in the
ladder approximation \cite{AGD75}, neglecting interference between
$a$ and $b$ channels. In this approximation, the problem reduces
to solving the Bethe-Salpeter equation in each channel.

This results in the renormalized scattering amplitudes
$\Gamma^{\left(i\right)}(\vec{q}_{1},\vec{q}_{2};\vec{q}_{3},\vec{q}_{4})$
for $i=a,b$. Here $\vec{q}_{3},\vec{q}_{4}$ and
$\vec{q}_{1},\vec{q}_{2}$ are magnon momenta before and after
scattering, respectively, and $\vec{q}_{1} + \vec{q}_{2} =
\vec{q}_{3} + \vec{q}_{4}$. The total energy of scattered magnons
was set to zero. This limit is compatible with our main goal to
describe the phase transition near $B_{c}$ when approaching the
phase boundary $T_{c}\left(B\right)$ from higher temperatures. At
$T\rightarrow T_{c}$, only the magnon states at
$\vec{q}\simeq\vec{Q}_{1,2}$ are occupied and the magnon spectrum
renormalization near the minima is important.

With given $\Gamma^{\left(a\right)}$ and $\Gamma^{\left(b\right)}$,
the complete set of two-particle scattering amplitudes was then
obtained by multiplying $\Gamma^{\left(a\right),\left(b\right)}$ by
products of four $\alpha_{q}$ and $\beta_{q}$ coefficients.  For
instance, a scattering process $(A_{\vec{q}_{3}},
B_{\vec{q}_{4}})\rightarrow(A_{\vec{q}_{1}},B_{\vec{q}_{2}})$ in the
channel $a$ is described by the amplitude
$\alpha_{q_{1}}\beta_{q_{2}}\alpha_{q_{3}}\beta_{q_{4}}
\Gamma^{\left(a\right)}(\vec{q}_{1},\vec{q}_{2};\vec{q}_{3},\vec{q}_{4})$.

The renormalization of low-energy magnons was found by treating
the magnon scattering effects in the Hartree-Fock approximation:

\begin{eqnarray}
H_{int}^{MF}=&&2\Gamma n\sum_{q}\left(  A_{q}^{+}A_{q}+B_{q}^{+}B_{q}\right)\nonumber\\
&&+2\Gamma^{\prime}\alpha\beta n\sum_{q}\left(  A_{q}^{+}B_{q}+B_{q}^{+}A_{q}\right),
\label{theory4}
\end{eqnarray}

\noindent
where $\alpha_{Q_{1}}^{2}=\alpha_{Q_{2}}^{2}\equiv\alpha^{2}$ and
$\beta_{Q_{1}}^{2}=\beta_{Q_{2}}^{2}\equiv\beta^{2}$. Taking into
account that $\alpha^{2}\gg\beta^{2}$, we keep here only the leading
contributions to energy parameters $\Gamma$ and $\Gamma^{\prime}$:

\begin{eqnarray}
\Gamma&&\simeq\alpha^{4}\Gamma^{\left(  a\right)  }\left(  \vec{Q}_{1},\vec
{Q}_{1};\vec{Q}_{1},\vec{Q}_{1}\right)  \nonumber\\&&=
\alpha^{4}\Gamma^{\left(  b\right)
}\left(  \vec{Q}_{2},\vec{Q}_{2};\vec{Q}_{2},\vec{Q}_{2}\right)
,\ \ \ \Gamma^{\prime}\simeq2\Gamma/\alpha^{2}\label{theory5}
\end{eqnarray}

\noindent and we obtained the estimate $\Gamma\simeq0.85J$.
According to eq.~\ref{theory4}, the chemical potential of magnons is
renormalized $\mu _{0}\rightarrow\mu_{eff}=\mu_{0}-2\Gamma n$, and the
low-energy magnons are hybridized due to the second term in
eq.~\ref{theory4}. This term shifts the bottom of the magnon band
slightly down and leads to a weak mass enhancement of low-energy
magnons. Both effects are proportional to $n^{2}$ and we omit them
since $n\ll1$ near $T_{c}$ for $\left( B_{c}-B\right) \ll B_{c}$.

For a given $B\lesssim B_{c}$ and with decreasing temperature the
magnon BEC occurs when the effective chemical potential
$\mu_{eff}$ vanishes \cite{Nikuni00}. Then $T_c(B)$ is determined
by
\begin{equation}
g\mu_{B}\left(B_{c}-B\right) =2\Gamma n\left( T_{c}\right)~.
\label{eq:gap}
\end{equation}
\noindent Here $n\left(T\right) =\sum_{q}f_{B}\left(E_{q}\right)$
with $f_{B}\left(E_{q}\right)$ being the Bose distribution function
taken at $\mu_{eff}=0$ and $E_{q}=E_{q}^{A}$ or $E_{q}=E_{q}^{B}$.
This means that for $T<T_{c}$ the magnon condensate develops
simultaneously at $\vec{q}=\vec{Q}_{1,2}$. It is worth emphasizing
that at $\mu_{eff}\rightarrow 0$ the distribution function $f_B(E)$
diverges as $T/E$ for $E\rightarrow 0$. Therefore, the low energy
3D-magnon spectrum, $E<E^{\ast}$, mainly contributes and drives the
BEC transition. The phase boundary can be calculated using
eq.~\ref{eq:gap}. It gives a very good description of the experimental
data near $B_{c}$ (see inset to Fig.~\ref{fig:phasediagram}), but
deviates strongly at lower fields, i.e., for $B_{c}-B>0.5$~T. This
indicates that the mean-field description of the magnon BEC is only
applicable in the close vicinity of $B_{c}$. The calculated boundary
is well described by eq.~\ref{eq:tcvsb} with a critical exponent
$\phi_{th}\simeq 1.5$ close to the predicted value $\phi_{\rm
  BEC}=3/2$ characteristic for 3D quadratic dispersion of low-energy
magnons \cite{Nikuni00,Nohadani04}.

We have presented experimental evidence that in \ccc the field
dependence of the critical temperature $T_c(B)\propto
(B_c-B)^{1/\phi}$ close to the critical field $B_c=8.51$~T is well
described with $\phi\simeq 1.5$. This is in very good agreement with
the exponent expected in the mean-field approximation. Together with
the observed opening of a spin gap above $B_c$ these findings support
the notion of a Bose-Einstein condensation of magnons in
Cs$_2$CuCl$_4$.

%
%
We are grateful to D. Ihle, B. Schmidt, M. Sigrist, P. Thalmeier, Y.
Tokiwa, and M. Vojta for stimulating discussions. V. Y.  acknowledges
financial support by the Deutsche Forschungsgemeinschaft.

%
%


\begin{thebibliography}{8.}
\bibitem{alphab2} The nuclear contributions were determined from a
  plot of $C_{tot}T^2$ vs $T^3$, assuming that $C_{tot}=C_{nuc} +
  C_{mag} + C_{ph}$. The nuclear contributions are due to the
  hyperfine interactions and can be written as
  $C_{nuc}=\alpha/T^2=(\alpha_Q + \alpha_Z)/T^2$, with the quadrupolar
  and the Zeeman contribution, respectively. In this way we obtained
  $\alpha_Q=33(11)$~$\mu$JK/mol and $\alpha_Z=13.8(3)
  B^2$~$\mu$JK/(mol~T$^2$). The latter value is in good agreement with
  the estimated value $\alpha_Z=11~\mu$JK/mol, using NMR-data
  \cite{NMR}.
\bibitem{mapping} This can be achieved through $S_{i}^{+}\rightarrow
  a_{i}$, $S_{i}^{-}\rightarrow a_{i}^+$, and
  $S_{i}^{z}=1/2-a_{i}^{+}a_{i}$, and a similar description for
  $b_{j}$.
\end{thebibliography}
\end{document}